\documentclass[12pt]{aastex}
\shorttitle{SMC WR}
\shortauthors{Massey, Olsen, \& Parker}

\begin{document}

\title{The Discovery of a Twelfth
Wolf-Rayet Star \\ in the Small Magellanic Cloud} 

\author{Philip Massey\altaffilmark{1}}

\affil{Lowell Observatory, 1400 W. Mars Hill Road, Flagstaff, AZ 86001}
\email{Phil.Massey@lowell.edu}

\author{K. A. G. Olsen\altaffilmark{1}}
\affil{Cerro Tololo Inter-American Observatory, National Optical Astronomy
Observatory, Casilla 603, La Serena, Chile}
\email{kolsen@noao.edu}

\and

\author{J. Wm. Parker\altaffilmark{1}}
\affil{Department of Space Studies, Southwest Research Institute, 1050
Walnut Street, Suite 400, Boulder, CO 80302}
\email{joel@boulder.swri.edu}

\altaffiltext{1}{
Visiting astronomer, Cerro Tololo Inter-American Observatory,
a division of the National Optical Astronomy Observatory, which is
operated by the Association of Universities for Research in Astronomy,
Inc., under cooperative agreement with the National Science Foundation.}

\begin{abstract}

We report the discovery of a relatively faint ($V=15.5$) early-type WN 
star in the SMC.  The line strength and width of He~II $\lambda 4686$ emission
is similar to that of the other SMC WNs, and the presense of 
N~V $\lambda 4603,19$ emission (coupled with the lack of N~III) suggests
this star is of spectral type WN3-4.5, and thus
is similar in type to the other SMC WRs.  Also like the other SMC WN stars,
an early-type absorption spectrum is weakly present. The absolute magnitude
is comparable to that of other (single) Galactic early-type
WNs. The star is located in the Hodge~53 OB association, which is
also the home of two other SMC WNs.   This star, which we designate SMC-WR12,
was actually
detected at a high significance level in an earlier interference-filter survey,
but the wrong star was observed as part of a spectroscopic followup, and this
case of mistaken identity resulted in its Wolf-Rayet nature not being
recognized until now.

\end{abstract}


\section{Introduction}

Wolf-Rayet stars (WRs)
are the evolved, He-burning descendents of the most massive
stars, and their strong emission lines allow them to serve as an important
tracer of the massive star content of nearby galaxies (see Massey 2003 for
a recent review).  Massey \& Duffy (2001) recently completed a survey
for Wolf-Rayet stars in the SMC, discovering two previously unknown ones,
bringing the total to 11 known.  Their study confirmed that there is
not a significant population of WR stars still to be found
in the SMC, although they expected that a few remained to be discovered.   
As they note, the number of WRs is thus
a factor of 3-4 lower in the SMC than in the LMC (normalized per unit
luminosity) despite the fact that the SMC and 
LMC have a comparable star-formation rate for massive stars.  
This is in accord with the suggestion
that at the lower metallicity of the SMC only the highest mass massive
stars possess sufficient stellar winds to evolve to the Wolf-Rayet phase.
Studies of coeval regions in the SMC, LMC, and Milky Way seem to confirm
this, as  the turn-off masses in clusters containing WRs suggest that
only stars with masses greater than $65\cal M_\odot$ may
become WRs in the SMC, while the lowest progenitor mass for a WR in the
LMC may be 30 $\cal M_\odot$  and $20\cal M_\odot$ in the
Milky Way (Massey, Waterhouse, \& DeGioia-Eastwood 2000;
Massey, DeGioia-Eastwood, \& Waterhouse 2001).

The authors and several additional
collaborators are engaged in a spectroscopic survey of hot, massive
stars in the Magellanic Clouds using the CTIO 4-m.  During a recent
run, we chanced across another previously unknown WR in the SMC.
Here we describe this interesting object.

\section{Observations and Reductions}
The data were taken on the CTIO 4-m Blanco telescope during a four
night observing run 18-21 December 2002 using the Hydra multi-object
fiber position (Barden \& Ingerson 1998).  The instrument consists of
138 fibers (300$\mu$m, which equals 2.0-arcsec in diameter) which can
be positioned within a 40 arcmin diameter field of view.  The fibers
``feed" a bench-mounted spectrograph, where we used grating KPGLD in
second-order and a BG-39 blocking filter and a 400-mm focal length
camera, behind which was a SITe 2096$\times$4096 (15$\mu$m pixels) CCD.
The chip was binned by 2 in the dispersion direction, resulting in a
dispersion of 0.45\AA\ pixel$^{-1}$ and a spectral resolution of 3.5
pixels (1.6\AA).  Our wavelength coverage extended from 3900\AA\ to
4950\AA.  The fiber-to-fiber sensitivity was removed by a combination of
exposures of the illuminated ``great white spot" each afternoon/morning
along with projector flat exposures taken at each telescope position
and fiber configuration.  The pixel-to-pixel variations of the CCD were
removed by means of a ``milk flat", an exposure through a diffuser screen
of the illuminated fiber ends. Wavelength calibration was by means of
a long He-Ne-Ar lamp exposure taken each afternoon, supplemented by
shorter exposures taken at each field.

The data for the WR star was obtained on the second night of the run
(19 Dec 2002).  A sequence of four exposures, each of 1200~s, was taken
of this field. The data were combined after extraction and processing.
The seeing conditions were described in the observing log as ``rotten".
At the time the Tololo seeing monitor was reporting the seeing as 2.4 arcsec,
eventually improving to 1.2 arcsec near the end of the exposures.

\section{Discussion: SMC-WR12}
\begin{figure}
\epsscale{0.9}
\plotone{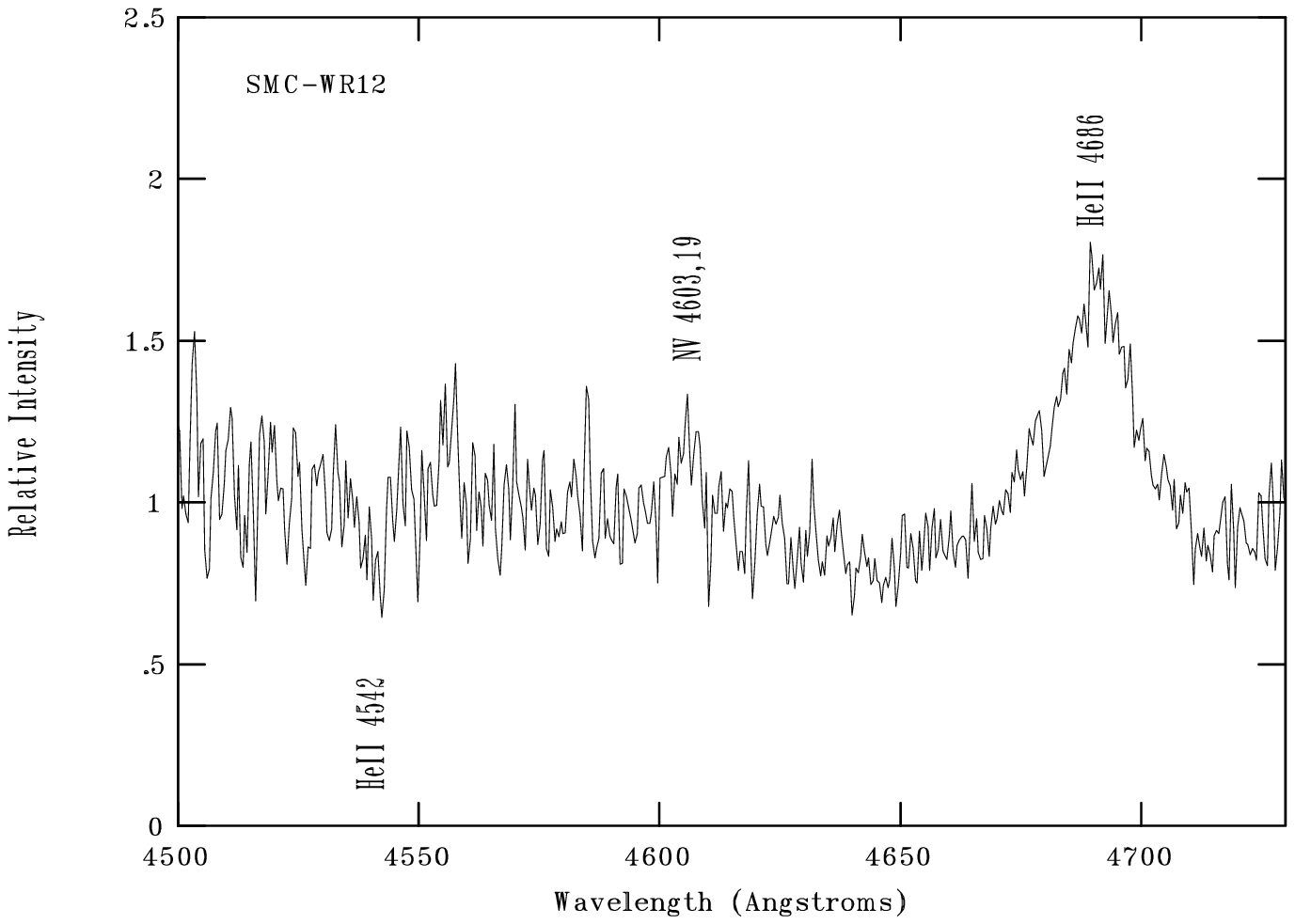}
\caption{\label{fig:spect} The spectrum of SMC WR-12 is shown. Strong, broad
emission lines 
of He~II $\lambda 4686$ and N~V $\lambda 4603, 19$ are visible, as well
as absorption at He~II $\lambda 4542$.}
\end{figure}

The spectrum of one of the targets in this field showed the characteristic
broad, strong emission features of a Wolf-Rayet star of 
the WN sequence (Fig.~\ref{fig:spect}).
He~II $\lambda 4686$ is visible with an equivalent width (EW)
of $-22$\AA\ and
a full-width at half-maximum of 21\AA.  The strength 
of He~II $\lambda4686$
argues that this star must be a WN-type Wolf-Rayet
rather than an Of-type star, as even the
most extreme Of-type stars known have EWs $>-10$\AA\ (Conti \& Leep 1974),
although
some SMC WNs stars do have EWs that overlap with Of stars 
(see 
Conti, Garmany, \& Massey 1989 and Conti \& Massey 1989).
The weak presence of N~V $\lambda 4603, 19$ emission 
also precludes the possibility that the star is an Of supergiant.
The lack of
N~III $\lambda 4634, 42$ emission then
makes the spectral class of type WN3-4.5. 
Unfortunately our spectrum is too noisy in the far blue to tell if 
NIV$\lambda 4058$ is
present or not, leading to the uncertainity in 
the subtype (Smith 1968, van der Hucht et al.\ 1981).  
In any event the spectral type
of this WN is ``early", similar to most of the other SMC WNs.
Absorption is clearly
visible at He~II $\lambda 4542$, although there is no sign of He~I
$\lambda 4471$ in our somewhat noisy spectrum. We would thus describe
the absorption spectrum as O3-O4. Although in general
WRs do not show any absorption features, nearly all of the SMC WNs do,
and the absorption features are mostly of the same O3-4 class (i.e., He~II).
Massey \& Duffy (2001) argue that the presence of absorption spectra in
the SMC WRs is still not well-understood: either it is due to the fact
that the stellar winds are weak (and hence one sees photospheric absorption)
or it suggests that most of the SMC WRs are binaries.  Recent radial
velocity studies by Foellmi, Moffat, \& Guerrero (2003) suggest that the
binary fraction of WRs is normal in the SMC.

We compare the spectral characteristics  and photometry to that of the other
SMC WRs in Table~1.
The values for this newly found WR star (which we are designating SMC-WR12
for consistency; see Massey \& Duffy 2001) are in fact in keeping with
those of the other SMC stars.  
The most notable thing about this WR star is how similar it is to the others
in
terms of all of its properties, although it is on the faint end of the
luminosity distribution of WRs in the SMC.  However, the absolute
visual magnitude (inferred by adopting $(B-V)_o=-0.32$, following Pyper 1966
and assuming a true distance modulus to the SMC of 18.9, following 
van den Bergh 2000) is quite normal for a (single) early-type WN in
the Milky Way (Conti \& Vacca 1990).

\begin{figure}
\epsscale{0.65}
\plotone{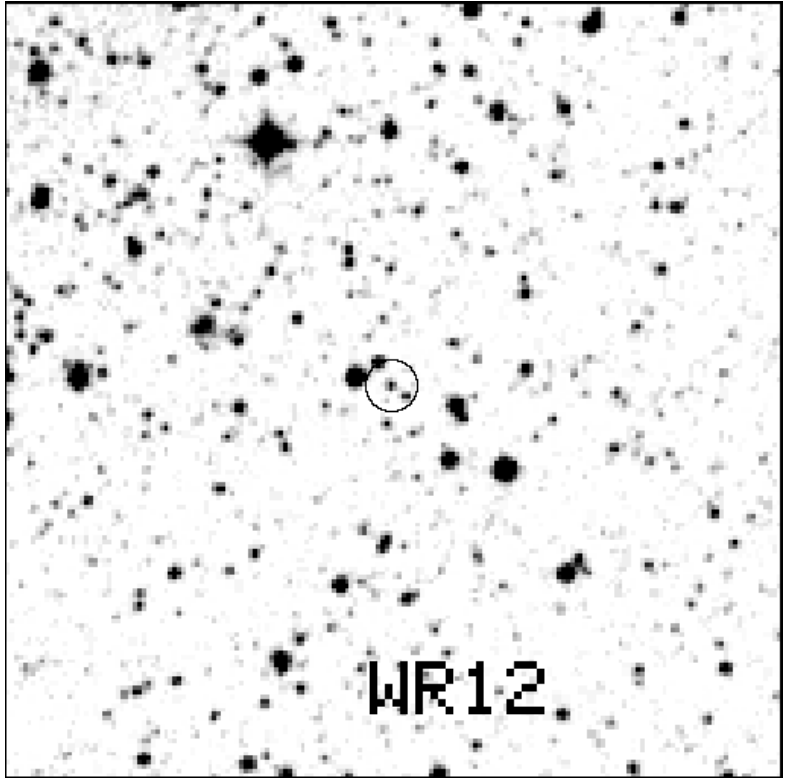}
\caption{\label{fig:fc}  Findging chart for SMC WR-12.  The chart
is 5 arcmin on a side and was made from the Digitized Sky Survey;
the circle marking the WR star is 20 arcsec in diameter.  North is up,
and east is to the left. Similar finding charts
for the other SMC WRs may be found in Massey \& Duffy (2001).
}
\end{figure}

SMC-WR12 was previously cataloged in the {\it UBVR} photometry of
Massey (2002) as SMC-054730.  
A finding chart is given in Fig.~\ref{fig:fc}.
The star is located in the Hodge~53 OB association (Hodge 1985), which
is also home to two of the other SMC WRs.  Thus one-quarter of the known
SMC WRs are found in this one rich association.  An investigation of
the {\it unevolved} stars in Hodge~53 was carried out by Massey et al.\ (2000),
who found that the most massive H-burning stars had (initial) masses
in the range of 50-80$\cal M_\odot$.  This would suggest that the progenitors
of the Hodge~53 Wolf-Rayet stars had masses of comparable or slightly
higher values, although the coevality of the region 
was considered ``questionable" as there 
were also evolved stars of 10-20$\cal M_\odot$ present.

We were naturally curious as to why this star was not detected on earlier
surveys.  The star is rather faint ($V=15.5$) to have been detected by 
the objective prism search of Azzopardi \& Breysacher (1979), who cataloged
the first eight WRs in Table~1. (Some had been known previous to their survey.)
In addition, the star's location in a relatively crowded region would create
confusion for objective prism studies.  Inspection of the working notes
for the Massey \& Duffy (2001) interference-filter imaging survey reveals
that SMC-WR12 {\it was} detected at a very high significance level ($7\sigma$),
with a magnitude difference (0.36~mag) consistent with real WRs.  
The star was observed spectroscopically as part of that program, but
the spectrum was that of a (foreground) 
G-type dwarf, and the first author incorrectly concluded
that there had been something wrong with the interference-filter photometry:
our notes say ``2 stars", indicated we thought that crowding had compromised
the photometry.  In retrospect, the wrong star must have been observed
spectroscopically.  A careful comparison of the telescope coordinates for
the old (October 2000) observations with that expected suggests that a
star about 5-10 arcsec south was observed instead, likely the star 7 arcsec
to the SW
shown on the finding chart. 
This is unfortunate, but consistent with Massey \& Duffy's
caution that they ``cannot preclude a WR star or two [from] having been
overlooked in our survey, particularly in crowded regions."

\acknowledgments
We are grateful for the generous allocation of observing time at CTIO, and
the (as usual) excellent support received from the mountain staff.
PM's role in this project was supported
by the National Science Foundation through grant AST0093060.
JWP's work was supported under NASA grant NAG5-9248.
PM also thank Alaine Duffy, whose excellent note-taking during the October 2000
observing run made it easy to trace the case of the mistaken identity of
the star actually observed.

\clearpage
\begin{deluxetable}{l l c c c c c c c c c c l}
\renewcommand{\arraystretch}{0.6}
\tabletypesize{\scriptsize}
\rotate
\tablewidth{0pc}
\tablenum{1}
\tablecolumns{13}
\tablecaption{Wolf-Rayet Stars in the SMC\tablenotemark{a}}
\tablehead{
\colhead{Star}
&\colhead{Other ID}
&\colhead{$\alpha_{2000}$}
&\colhead{$\delta_{2000}$}
&\colhead{OB/HII\tablenotemark{b}}
&\colhead{Spectral Type}
&\colhead{Abs.}
&\colhead{$V$}
&\colhead{$B-V$}
&\colhead{$M_V$}
&\multicolumn{2}{c}{He~II or C~III\tablenotemark{c}}
&\colhead{Comment}  \\ \cline{11-12}
&&&&&&&&&&\colhead{EW (\AA)}
&\colhead{FWHM(\AA)} \\
}
\startdata
SMC-WR1 & AV~2a & 00 43 42.23 &  $-$73 28 54.9 & no & WN3+abs  &O3-4 & 15.14 & $
-0.04$   &$-4.6$  &  $-$28 & 21 & Weak abs. \\
SMC-WR2 & AV~39a& 00 48 30.81 &  $-$73 15 45.1 & near h15 &WN4.5+abs&O5: & 14.23
 & $-$0.15   &$-5.2$  &  $-$15 & 12\\
SMC-WR3 & AV~60a& 00 49 59.33 &  $-$73 22 13.6 & near h17 &WN3+abs &\nodata&14.48
& $-$0.10  &$-5.1$  &  $-$53 & 26 & Very weak abs. \\
SMC-WR4 & AV~81, Sk~41& 00 50 43.41 &  $-$73 27 05.1 &h21&  WN6p&\nodata &13.35&
 $-$0.16&$-6.2$ & $-$45 & 15 & N~V abs?\\
SMC-WR5 & HD~5980 & 00 59 26.60 &  $-$72 09 53.5&NGC~346=h45 & WN5&\nodata &11.08
& +0.03 & $-$8.9 &  $-$85 & 18 \\
SMC-WR6 & AV 332, Sk 108 & 01 03 25.20 &  $-$72 06 43.6 &NGC~371(e76)=h53 
& WN3+abs &O7& 12.30
&$-$0.15 & $-7.1$&  $-$8 & 28 \\
SMC-WR7 & AV~336a & 01 03 35.94 &   $-$72 03 21.5 & NGC~371(e76)=h53&  WN2+abs & O6&12.93&$-$
0.05 &$-6.8$ &  $-$16 & 27 \\
SMC-WR8 & Sk 188 & 01 31 04.22 &  $-$73 25 03.9 & NGC~602c=h69 & WO4+abs &O4~V&
12.81 &$-$0.14 &$-6.6$ &  $-$76 & 71 \\
SMC-WR9 & Morgan et al.\ & 00 54 32.17 &  $-$72 44 35.6 & no & WN3+abs &O3-4&15.23
&$-$0.13&$-4.3$ &$-22$ & 24\\
SMC-WR10 &  & 00 45 28.78 &  $-$73 04 45.2 & NGC~249(e12)&WN3+abs & O3-4&15.76:&
$-$0.08: &$-3.6$ &  $-$24 & 23 & Strong neb.\\
SMC-WR11 &  & 00 52 07.36 &  $-$72 35 37.4 & no & WN3+abs & O3-4&14.97&+0.18&$-5
.5$ &  $-$14 & 25 & \\
SMC-WR12 & SMC-054730 & 01 02 52.07 &  $-$72 06 52.6 & NGC~371(e76)=h53 & WN3-4.5+abs & O3-4 & 15.46 & $-$0.15 & $-4.0$ & $-$22 & 21 & \\

\enddata
\tablenotetext{a}{This table is an updated version of Table~I in Massey \& Duffy 2001.}
\tablenotetext{b}{OB associations designation (h) are from Hodge (1985), and
emission-line regions are from the Hodge \& Wright (1977) atlas.}
\tablenotetext{c}{The equivalent width (EW) and full-width-at-half-maximum
(FWHM)
of He~II~$\lambda4686$ is given for
the WNs; that of C~III~$\lambda4650$ is given for the WO4 star.}
\end{deluxetable}

\end{document}